\documentstyle[12pt]{article}
\newcommand{\tra}{{\rm Tr}({\cal H})}
\newcommand{\be}{\begin{equation}}
\newcommand{\ee}{\end{equation}}
\newcommand{\sli}{S_{\rm lin}}
\newcommand{\tet}{\overline{T}_t}
\newcommand{\sui}{\sum\limits_{i=1}}
\newcommand{\suj}{\sum\limits_{j=1}}
\newcommand{\trp}{{\rm Tr}({\cal H})_+}
\newcommand{\tiso}{{\rm Tr}({\cal H})_{\rm iso}}
\newcommand{\hsh}{{\rm HS}({\cal H})}
\newcommand{\mkn}{{\cal M}_{kn}}
\newcommand{\hikn}{\tilde{\cal H}_{kn}}
\newcommand{\bha}{{\rm B}({\cal H})}
\newcommand{\mnn}{M_{n\times n}}
\newcommand{\hac}{{\cal H}\otimes{\bf C}^n}
\newcommand{\hsa}{{\rm HS}(\hac )}
\newcommand{\trhp}{{\rm Tr}(\hac )_+}
\newcommand{\kom}{{\rm K}({\cal H})}
\newcommand{\pef}{\hat{P}\phi}
\newcommand{\pkn}{\overline{P}_{kn}}
\newcommand{\bikn}{{\rm B}(\hikn )}
\newcommand{\mrk}{M_{r(k)\times r(k)}}
\newcommand{\beh}{{\rm B}(E_{kn}{\cal H})}
\newcommand{\pna}{\pkn(A)}
\newcommand{\nkn}{{\cal N}_{kn}}
\newcommand{\unn}{{\cal U}(\nkn )}
\newcommand{\cun}{\int\limits_{\unn}d\mu(U)}
\newcommand{\crk}{\int\limits_{U(r(k))}d\mu(U)}

\begin{document}
\title{Structure of the Algebra of Effective Observables in Quantum Mechanics}
\author{R. Olkiewicz\thanks{Supported by the KBN research grant No 2 P03B
086 16}\\Institute of Theoretical
Physics, University of Wroc{\l}aw\\ PL-50-204 Wroc{\l}aw, Poland}
\date{ }
\maketitle
\begin{abstract}
A subclass of dynamical semigroups induced by the interaction of a 
quantum system with an environment
is introduced. Such semigroups lead to the selection of a stable subalgebra 
of effective observables. The structure of this subalgebra is completely
determined.
\end{abstract}
\newpage
\noindent
{\bf 1 Introduction}\\[3mm]
One of the fundamental principles of quantum mechanics 
is the superposition principle which guarantees
that any superposition of two distinct pure states 
is again a legitimate pure state. As an immediate
consequence of this and the postulate that proportional 
vectors describe the same quantum state we obtain
that pure states are in one-to-one correspondence with 
one-dimensional subspaces of a Hilbert space
$\cal H$. This is usually taken as a basic ingredient of 
a mathematical description of a system, which
ensures genuine quantum behavior of that system. 
Alternatively, we may say that physical quantities are
in one-to-one correspondence with self-adjoint operators 
on $\cal H$. Since, without loss of generality,
we may restrict to bounded operators so it implies that 
the von Neumann algebra $\cal A$ generated by 
observables equals to B$({\cal H})$, the algebra 
of all bounded operators. However, it is evident
that some superpositions of quantum pure states do not 
take place in the real world. Well known examples
of such a phenomenon encompass the absence of 
superpositions of states with different electric charge or 
with integer and half-integer spin. This fact led to the introduction in 
1952 of superselection rules [1], which axiomatically 
exclude certain superpositions from being 
observable. For a review of this subject see a 
recent paper by Wightman [2]. It follows that the connection
between quantum states and rays in $\cal H$ should 
be changed to: every pure state is represented by a
one-dimensional subspace of $\cal H$, but not every such 
a subspace represents a quantum state. Such a
postulate has an immediate consequence for the 
algebra $\cal A$, since now the commutant ${\cal A}'$, 
which consists of superselection operators, is non-trivial. 

Further, in 1960, Jauch [3] introduced a condition 
that there should exist at least one
complete set of commuting observables in $\cal A$, 
which expressed more generally states that $\cal A$
should contain a maximal Abelian subalgebra. 
It implies that all superselection operators
belong to $\cal A$ or, equivalently, that the 
center $\cal Z$ of $\cal A$ equals to ${\cal A}'$. Clearly,
all superselection operators commute with each other 
since  ${\cal A}'$ is Abelian in this case. Therefore,
the existence of superselection rules makes the 
center $\cal Z$ non-trivial yielding a decomposition
of Hilbert space $\cal H$ into coherent subspaces. 
In the discrete case it was concisely written by
Wan [4] as follows:\\
{\it Let $\cal S$ denote the set of all pure states. 
Then $\cal H$ may be decomposed into a direct sum of 
mutually orthogonal subspaces ${\cal H}_n$ such that 
${\cal S}\,=\,\bigcup_n{\bf C}P({\cal H}_n)$, where
${\bf C}P({\cal H}_n)$ denotes the projective space 
over ${\cal H}_n$. There is no further decomposition
of ${\cal H}_n$.}\\
It is worth noting
that a superposition $\lambda|v>+\lambda'|w>$, 
$|\lambda|^2+|\lambda'|^2=1$, of vectors from different
coherent subspaces is empirically indistinguishable from 
the mixture $|\lambda|^2P_v+|\lambda'|^2P_w$,
where $P_v=\,|v><v|$ and $P_w=\,|w><w|$.
As a consequence, the algebra $\cal A$ consists of 
all operators $A\in{\rm B}({\cal H})$ such that
$\sum_nP_nAP_n=\,A$, where $P_n$ denotes the orthogonal 
projector onto ${\cal H}_n$. 

A more general situation can also occur. When we 
drop Jauch's hypothesis we obtain that in principle
${\cal A}'$ has only a partial overlapping with 
$\cal A$. Hence, there are non-commuting superselection
operators since ${\cal A}'$ cannot be Abelian now. 
It means that $\cal A$, when restricted to a coherent 
subspace ${\cal H}_n$, is still smaller than 
B$({\cal H}_n)$. Therefore, some different and non-proportional
vectors from ${\cal H}_n$ may still determine the same 
quantum state. Such a possibility was explicitly
acknowledged by Messiah and Greenberg in 1964 [5], 
who introduced the term generalized ray for the set of
such vectors. In such a case the lack of knowledge 
of the state vector is greater than in ordinary quantum
mechanics. A generalized ray is represented by 
an $r$-dimensional sphere, $r$ being the dimension of an 
irreducible subspace of commuting physical observables. 
This inevitably puts additional constraints on
the structure of algebra $\cal A$.  We encounter such 
a situation when, for example, we want to study
symmetry transformations called supersymmetry [6], 
which leave all the observables invariant. Let us
recall that a unitary operator is a supersymmetry if 
it is not proportional to the identity operator
and commutes with the set of all observables. Clearly, 
they and the identity form a unitary group of
${\cal A}'$, so-called gauge group.

Superselection rules was a useful postulate, but 
the question about an explanation of its appearance arose.
It should be pointed out here that it is not a 
logical necessity of quantum theory. In 1982 Zurek [7]
proposed a program of environment-induced superselection rules. He
showed that when a quantum system is open, interacting 
with an environment, superselection rules do not need to
be postulated. They arise naturally as a result 
of the decoherence process, which effectively destroys 
superpositions between macroscopically different 
states with respect to a local observer, so that the system
appears to be in one or the other of those states. 
By the term ``destroys superposition'' we understand that
the off-diagonal elements of the superposition are 
unavailable with respect to a specific set of observations.
The idea was further developed in [8,9,10]. 

In order to study decoherence, the analysis of the 
evolution of the reduced density matrix obtained by
tracing out the environment variables is the most 
convenient strategy. For a large class of interesting
physical phenomena the evolution of the reduced density 
matrix can be described by a dynamical semigroup,
whose generator is given by a Markovian master equation.
The loss of quantum coherence in the Markovian regime 
was established in a number of open systems [11,12]
giving a clear evidence of environment-induced 
superselection rules. In a recent paper [13] a thorough
mathematical analysis of the superselection structure induced by a dynamical 
semigroup which is also contractive in the operator norm was presented.
It was achieved by the use of the isometric-sweeping 
decomposition, which singles out a subspace of density
matrices, on which the semigroup acts in a reversible, 
unitary way, and sweeps out the rest of statistical
states. The dual space of the isometric part of density matrices
is a von Neumann algebra $\cal M$, which we call the algebra of effective
observables. This algebra is stable with respect 
to the process of decoherence, i.e. its elements
evolve in a unitary way according to Schr\"odinger dynamics 
in the Heisenberg picture.
Other elements of B$({\cal H})$ decay in time
to elements of $\cal M$. Therefore, when decoherence 
happens almost instantaneously, then $\cal M$ represents
physical observables of the quantum system. The 
purpose of this paper is to describe the structure of $\cal M$.

The paper is organized as follows. In sec. 2 we introduced 
the notion of an environment-induced semigroup
and discuss its properties. In sec. 3 we briefly 
recall some basic facts concerning superselection rules
induced by the interaction with an environment. Finally, 
in sec. 4, we describe the structure of the
algebra of effective observables.\\[4mm]
{\bf 2 Environment-induced semigroups}\\[3mm]
The irreversible behavior of the evolution of 
quantum statistical states (density matrices) 
is the main consequence of the assumption that
they interact with their environments. As was 
mentioned in Introduction we restrict our considerations to the
Markovian regime, and thus assume that the evolution 
of the reduced density matrix is given by a dynamical
semigroup $T_t$. By a dynamical semigroup one usually 
means a strongly continuous semigroup of completely
positive trace preserving and contractive 
operators acting on the Banach space of trace class operators 
$\tra$ [14]. However, since the semigroup $T_t$ is to 
describe a measurement-like
interaction with the environment, the statistical 
entropy $S(\rho)\,=\,-{\rm tr}\rho\log\rho$
of an evolving density matrix $\rho$ should not decrease. Here, by tr we 
denote the usual trace on $\tra$. For a measurement it follows from the 
following argument. Suppose that the properties of a quantum system are 
specified by probabilities $\{p_i\}$ for the outcomes of the measurement of 
a discrete observable $A$. Therefore, the state of such a system is a mixed 
state and reads $\rho\,=\,\sum_ip_iP_i$, where $P_i\,=\,|e_i><e_i|$ and 
$|e_i>$ are the corresponding eigenvectors of $A$. The statistical entropy 
of $\rho$ is a measure of our ignorance of the actual result of the 
measurement of $A$. Suppose further that we perform a measurement of another 
discrete observable $B$. According to the von Neumann projection postulate 
the state of the system changes to \be \rho\to\rho '\;=\;\sum\limits_j 
Q_j\rho Q_j\ee 
where $Q_j=\,|f_j><f_j|$, $|f_j>$ being eigenvectors of $B$. 
Therefore $\rho '=\,\sum_jp_j'Q_j$, where 
$p_j'=\,\sum_ip_i{\rm tr}(Q_jP_i)$. 
Because coefficients tr$(Q_jP_i)$ form a doubly stochastic 
matrix so $S(\rho ')\geq S(\rho)$. For a more general discussion of the 
entropy increase during the interaction with the environment see [15].

However, the concept of dynamical semigroup as defined above is 
too general to ensure the increase of the
statistical entropy, as the following simple 
example shows. Consider an operator $L$ defined by
$$L\rho\;=\;A\rho A^*\;-\;\frac{1}{2}\{A^*A,\,\rho\}$$
where $\{\cdot,\cdot\}$ stands for the anticommutator and 
$$A\;=\;\left(\begin{array}{cc} 0&0\\1&0\end{array}\right)$$
Clearly, $L$ generates a dynamical semigroup $T_t$ 
on $2\times 2$ complex matrices. By direct calculations
we obtain that the evolution of the one-dimensional projector 
$$P\;=\;\left(\begin{array}{cc} 1&0\\0&0\end{array}\right)$$
is given by $T_tP\,=\,e^{-t}P\,+\,(1\,-\,e^{-t})P^{\bot}$, 
where $P^{\bot}\,=\,I\,-\,P$ 
and $I$ denotes the identity matrix. Hence the statistical
entropy increases to its maximal value when time approaches $\log 2$, 
and then decreases to zero. Although such semigroups may also play some role
in theoretical investigations of quantum open systems, they will be
excluded from our current considerations. Therefore, 
we impose on $T_t$ an additional assumption, namely that $T_t$ is also 
contractive in the operator norm $\|\cdot\|_{\infty}$, and we call it {\bf 
environment-induced semigroup}. For such a semigroup it follows that a 
maximal eigenvalue of a density matrix cannot increase during the evolution. 
Moreover, the following properties can be derived. First, notice that the 
linear entropy $\sli(\rho)\,=\,{\rm tr}(\rho\,-\,\rho^2)$, being a linear 
approximation of $S$ since 
$\log\rho\,=\,\log(I\,-\,(I\,-\,\rho))\,=\,\rho\,-\,I\,+...$, does not 
decrease. Indeed, for $t_1\geq t_2$
$$\sli(T_{t_1}\rho)\;-\;\sli(T_{t_2}\rho)\;=
\;\|T_{t_2}\rho\|_2^2\;-\;\|T_{t_1-t_2}(T_{t_2}\rho)\|_2^2\;\geq\;0$$
since, by Lemma 4 in [13], $T_t$ is also contractive in the Hilbert-Schmidt 
norm $\|\cdot\|_2$. The statistical entropy does not decrease, either. For 
finite dimensional quantum systems it follows from the following argument. 
For a totally mixed state $\rho_0=\,\frac{I}{{\rm tr}I}$, all eigenvalues 
of $T_t(\rho_0)$ are not greater than $\frac{1}{{\rm tr}I}$ and their sum is 
equal to 1. Hence $T_t(\rho_0)\,=\,\rho_0$ and so $T_t(I)\,=\,I$ for all 
$t\geq 0$. Thus the function $t\to S(T_t\rho)$ is non-decreasing for any 
density matrix $\rho$. We show that
this property also holds in the infinite dimensional 
case. \\
{\bf Proposition 2.1} {\it Suppose $T_t$ is an environment-induced 
semigroup. Then $S(T_t\rho)\geq S(\rho)$ for any density matrix} 
$\rho$.\\
{\bf Remark}. Since $S$ takes values in $[0,\,\infty]$ so the above 
inequality means that if $S(T_t\rho)<\infty$, then also $S(\rho)$ is finite 
and not greater than $S(T_t\rho)$.\\
{\bf Proof:} Let $\rho$ be a density 
matrix. Then $\rho\,=\,\sum_{i=1}p_iP_i$, $p_i>0$, and 
$T_t\rho\,=\,\sum_{j=1}q_jQ_j$, $q_j>0$, where $\{P_i\}(\{Q_j\})$ are 
spectral projectors of $\rho\;(T_t\rho)$ respectively. Clearly, all of them 
are finite dimensional. Projectors corresponding to zero eigenvalue we 
denote by$P_0$ and $Q_0$ respectively. Then
$$q_j\;=\;\frac{{\rm 
tr}(Q_jT_t\rho)}{{\rm tr}Q_j}\;=\;\sui 
p_i\alpha_j(i),\;\mbox{where}\;\alpha_j(i)\;=\;\frac{{\rm 
tr}(Q_jT_tP_i)}{{\rm tr}Q_j}$$
By Lemma 4 in [13], $T_t$ has a normal 
extension to a contractive semigroup $\tet$ on B$({\cal H})$. Suppose 
$\{E_n\}$ is a sequence of mutually orthogonal one-dimensional projectors 
such that $\sum_nE_n=\,I$.Then 
$$\tet(I)\;=\;\lim\limits_{n\to\infty}T_t(\sum\limits_{k=1}^nE_k)\:\leq\:I$$ 
Therefore $$I\:\geq\:\tet(I)\;=\;\tet(P_0\:+\:\sui 
P_i)\;=\;\tet(P_0)\;+\;\sui T_tP_i$$
and so $\sum_{i=1}\alpha_j(i)\leq 1$. 
Let us define $\alpha_j(0)\,=\,1\,-\,\sum_{i=1}\alpha_j(i)$. Then we 
have $q_j\,=\,\sum_{i=0}p_i\alpha_j(i)$, where $p_0=\,0$ and 
$\sum_{i=0}\alpha_j(i)\,=\,1$. Because function $x\to x\log x$ is convex and 
continuous with $f(0)\,=\,0$, so
$$q_j\log q_j\:\leq\:\sui\alpha_j(i)p_i\log p_i$$
Assume now that $S(T_t\rho)$ is finite. It means that $-T_t\rho\log 
T_t\rho$ is a positive and trace class operator. 
Therefore
$$S(T_t\rho)\;=\;-{\rm tr}(\suj (q_j\log q_j)Q_j)\;=\;-\suj q_j\log 
q_j{\rm tr}Q_j\:\geq$$
$$\sum\limits_{i,j=1}(-p_i\log p_i){\rm 
tr}(Q_jT_tP_i)$$ 
Because $p_iP_i\leq\rho$ so $T_t(p_iP_i)\leq T_t\rho$ and 
hence tr$(Q_0T_tP_i)\,=\,0$ for all $i\geq 
1$.Therefore
$$S(T_t\rho)\:\geq\:\sui (-p_i\log p_i)[{\rm 
tr}(Q_0T_tP_i)\:+\:\suj{\rm tr}(Q_jT_tP_i)]$$
$$=\;\sui (-p_i\log p_i){\rm 
tr}(T_tP_i)\;=\;S(\rho)$$
since $T_t$ is trace preserving. $\Box$\\
Having 
discussed the properties of 
environment-induced semigroups, we now turn to a 
condition which guarantees that a dynamical semigroup $T_t$ is also 
contractive in the operator norm. In order to avoid domain difficulties we 
restrict ourselves to a uniformly continuous semigroup. Then its generator 
$L:\tra\to\tra$ has the following standard form
\be 
L\rho\;=\;-i[H,\,\rho]\;+\;\suj V_j\rho V_j^*\;-\;\frac{1}{2}\{\suj 
V_j^*V_j,\,\rho\}\ee
where $H\,=\,H^*\in{\rm B}({\cal H})$, $V_j\in{\rm 
B}({\cal H})$ and $\lim_n\sum_{j=1}^nV_j^*V_j\,=\,V$ in the strong 
topology.\\
{\bf Proposition 2.2} {\it A uniformly continuous dynamical 
semigroup $T_t$ is contractive in the operator norm if and only if} 
$\sum_jV_jV_j^*\leq\sum_jV_j^*V_j$.\\
{\bf Proof:} $\Leftarrow$ Suppose 
$\sum_jV_jV_j^*\leq V$. Then it converges strongly to some positive 
operator in B$({\cal H})$ and so $L$ extends to a bounded operator on 
B$({\cal H})$. It is clear that such an extension is a complete dissipation 
and so $T_t$ is contractive in the operator norm.\\
$\Rightarrow$ Suppose 
that $\|T_t\phi\|_{\infty}\leq\,\|\phi\|_{\infty}$ for all $\phi\in\tra$. 
Then $T_t$ extends to a contractive semigroup $\tet$ on K$({\cal H})$, the 
space of compact operators. Clearly, $\tet$ is strongly continuous with 
$\tra\subset D(\overline{L})$, where $\overline{L}$ denotes the generator of 
$\tet$. In order to show that $\sum_jV_jV_j^*\leq\sum_jV_j^*V_j$ it suffices 
to check that for any one-dimensional projector $P$ the following inequality 
holds $${\rm tr}P\suj V_jV_j^*\;\leq\;{\rm tr}P\suj V_j^*V_j$$
Let us fix 
projector $P$. Using the decomposition ${\cal H}\,=\,P{\cal H}\oplus 
P^{\bot}{\cal H}$ each $V_j$ can be written as
$$V_j\;=\;\left(\begin{array}{cc} 
a_j&w_j^*\\v_j&A_j\end{array}\right)$$
where $a_j\in{\bf C}$, $v_j,\,w_j\in 
P^{\bot}{\cal H}$ and $A_j\in{\rm B}(P^{\bot}{\cal H})$. In consequence, the 
above inequality is equivalent to $\sum_j\|w_j\|^2\leq\sum_j\|v_j\|^2$. 
Suppose now $\phi\,=\,P\,+\,E$, where $E$ is a finite dimensional 
subprojector of $P^{\bot}$. Clearly, $\phi\in D(\overline{L})$. 
Moreover, since K$({\cal H})^*\,=\,\tra$, $P$ is a normalized tangent 
functional to $\phi$ and so, by the assumption and Hille-Yosida theorem, 
tr$\overline{L}(\phi)P\leq\,0$ or, equivalently, $${\rm tr}P(\suj V_j\phi 
V_j^*)\;\leq\;{\rm tr}P(\suj V_j^*V_j)$$
Because
$${\rm tr}PV_j\phi 
V_j^*\;=\;|a_j|^2\;+\;<w_j,\,Ew_j>$$ and$${\rm tr}P(\suj V_j^*V_j)\;=\;\suj 
|a_j|^2\;+\;\suj\|v_j\|^2$$
we obtain that 
$\sum_j\|Ew_j\|^2\leq\sum_j\|v_j\|^2$. Taking the supremum over $E$ ends the 
proof. $\Box$\\
[4mm]{\bf 3 Environment-induced superselection rules}\\[3mm]
In this section we briefly presents some basic facts concerning 
the superselection structure induced by the interaction with an environment. 
Clearly, there is a difference between them and the traditional 
superselection rules, which are said to operate between subspaces of a 
Hilbert space if the phase factors between vectors belonging to two distinct 
subspaces are unobservable. In the case of environment-induced 
superselection rules, phase coherence between vectors from some preferred 
set of pure states is being continuously destroyed by the interaction.

Suppose $\hat{P}$ is a linear, bounded and 
positive operator on $\tra$ such that $\hat{P}^2=\,\hat{P}$
and tr$\hat{P}\phi\leq\mbox{tr}\phi$ for all $\phi\in\trp$, 
the cone of positive elements in $\tra$.
We call such an operator the {\bf projection operator} 
(when, in addition, $\hat{P}$ preserves the trace,
it is usually called the Zwanzig projection). Then space $\tra$ splits into 
two linearly independent and closed subspaces 
$\hat{P}\tra$ and $(\mbox{id}-\hat{P})
\tra$. We start with the following general definition, see ref. 13.\\
{\bf Definition 3.1} {\it We say that the semigroup 
$T_t$ induces a weak superselection structure on $\tra$ if\\
a) there exists a projection operator $\hat{P}$ such that}
\be T_t:\mbox{im}\hat{P}\to\mbox{im}\hat{P},
\quad T_t|_{{\rm im}\hat{P}}\;=\;U_t\cdot U_t^* \ee
{\it where $U_t$ is a strongly continuous group of unitary operators,\\
b)} \be 
\lim\limits_{t\to\infty}|{\rm tr}AT_t\phi\:-\:{\rm 
tr}A\hat{P}(T_t\phi)|\;=\;0 \ee
{\it holds for all $\phi\in\tra$ and any $A$ 
from some $^*$-algebra $\cal B$, which is strongly dense in $\bha$.\\
$T_t$ 
induces a strong superselection structure if a) holds together with\\
b')} 
\be 
\lim\limits_{t\to\infty}\|T_t\phi\:-\:
\hat{P}(T_t\phi)\|_1\;=\;0\quad\forall\phi\in\tra \ee
{\it where $\|\cdot\|_1$ is the trace norm. A weak(strong) superselection 
structure is said to be non-trivial if $\hat{P}\neq\mbox{id}$, conservative, 
if ${\rm tr}\hat{P}\phi\,=\,{\rm tr}\phi$ for all} $\phi\in\tra$.\\
It follows that environment-induced semigroups 
always induce (possibly trivial 
as they can be of purely unitary type) a superselection structure.\\
{\bf 
Theorem 3.2} [13] {\it Suppose $T_t$ is an environment-induced semigroup. 
Then $T_t$ induces a weak superselection structure. If moreover, $T_t$ is 
relatively compact in the strong operator topology, then it induces a strong 
superselection structure.}\\
{\bf Sketch of proof:} Let $K\subset\hsh$ be a  
subspace of the Hilbert space of Hilbert-Schmidt operators given 
by
$$K\;=\;\{x\in\hsh:\;\:\|T_tx\|_2\:=\:\|T_t^*x\|_2\:=\:\|x\|_2\;\forall 
t\geq 0\}$$
where $T_t^*$ denotes the conjugate with respect to the scalar 
product in $\hsh$. Let $\hat{P}:\hsh\to\hsh$ be the orthogonal projector 
onto $K$. It turns out that $\hat{P}$ maps trace class operators into trace 
class operators and tr$\hat{P}\phi\leq{\rm tr}\phi$ for any $\phi\in\trp$. 
Therefore, $\hat{P}$ induces a splitting $\tra\,=\,\tiso\oplus\tra_{\rm s}$ 
to the isometric and sweeping parts which fulfill the conditions from 
Definition 3.1. $\Box$\\
Let $\cal M$ be a von Neumann algebra having $\tiso$ 
as its predual space, that is ${\cal M}=\,{\rm im}\hat{P}^*$, where 
$\hat{P}^*:\bha\to\bha$ is the conjugate projector. We call it algebra 
of effective observables. The action of the dual semigroup 
$T_t^*:\bha\to\bha$, when restricted to $\cal M$, is given 
by a unitary 
group of automorphisms. In the next section we describe the structure of 
$\cal M$.\\[4mm]
{\bf 4 Algebra of effective observables}\\[3mm]
At first we 
show the following theorem.\\
{\bf Theorem 4.1} {\it Suppose $K\neq 0$. 
Then\\
a) $\hat{P}$ is completely positive i.e. $\forall 
n\;\hat{P}\otimes{\rm id}_{n\times n}:\tra\otimes\mnn\to\tra\otimes\mnn$ 
maps positive operators on $\hac$ into positive ones. Here $M_{n\times 
n}$denotes the algebra of $n\times n$ complex matrices.\\
b) 
$\|\hat{P}\|_{\infty ,\infty}=\,1$, $\|\hat{P}\|_{1,1}=\,1$.\\
c) $\hat{P}$ 
can be extended to a normal norm one projection $\overline{P}:\bha\to\bha$ 
onto the von Neumann algebra} $\cal M$.\\
{\bf Proof:} a) Let 
$K_n:=\,K\otimes\mnn\subset\hsh\otimes\mnn\,=\,\hsa$. If 
$\tilde{x},\tilde{y}\in K_n$, then also $\tilde{x}\tilde{y}\in K_n$ and 
$\tilde{x}^*\in K_n$, since $K$ is a $^*$-algebra. Suppose 
$\tilde{x}=\tilde{x}^*\in K_n$. Because $\tilde{x}$ is a Hilbert-Schmidt 
operator on $\hac$, so $\tilde{x}=\,\sum_ia_i\tilde{e}_i$, $a_i\in{\bf 
R}\setminus\{0\}$. Since $K_n$ is closed we obtain that $\tilde{e}_i\in K_n$ 
for all $i$. Suppose now that $\tilde{\phi}\in\trhp$. Then 
$\tilde{\phi}=\tilde{\phi}_1+\tilde{\phi}_2$, where $\tilde{\phi}_1\in K_n$ 
and $\tilde{\phi}_2\in K_n^{\bot}=\,K^{\bot}\otimes\mnn$. Because 
$\tilde{\phi}_1$ is hermitian, so $\tilde{\phi}_1=\,\sum_ib_i\tilde{e}_i$, 
$b_i\neq 0$ and $\tilde{e}_i\in K_n$. Hence $\tilde{\rm 
tr}\tilde{e}_i\tilde{\phi}_2=\,0$ for any $i$, what implies 
that $b_i=\,\tilde{\rm tr}\tilde{e}_i\tilde{\phi}/\tilde{\rm 
tr}\tilde{e}_i$. Therefore $\tilde{\phi}_1\geq 0$ and 
$$\tilde{\rm 
tr}\tilde{\phi}_1\;=\;\sum_i\tilde{\rm 
tr}\tilde{e}_i\tilde{\phi}\:\leq\:\tilde{\rm tr}\tilde{\phi}$$
Hence 
$\tilde{\phi}_1\in\trhp$ and so $\hat{P}^{(n)}:\trhp\to\trhp$, where 
$\hat{P}^{(n)}$ denotes the orthogonal projection in $\hsa$ onto $K_n$. 
However, $\hat{P}^{(n)}\,=\,\hat{P}\otimes{\rm id}_{n\times n}$, what 
implies that $\hat{P}$ is n-positive.
Because n was arbitrary, the assertion 
follows.\\
b) By point a), $\|\pef\|_1\leq\,\|\phi\|_1$ for all 
$\phi\in\trp$. Hence $\hat{P}$ is a bounded operator on $\tra$ with 
$\|\hat{P}\|_{1,1}\leq\,2$. Because $\tra\subset\kom$ and $\kom^*=\,\tra$, 
so for any $\phi\in\tra$, 
$$\|\pef\|_{\infty}\;=\;{\rm tr}(\pef 
)\psi\;=\;{\rm tr}\phi\hat{P}(\psi)$$
for some $\psi\in\tra$ with 
$\|\psi\|_1=\,1$. 
Hence
$$\|\pef\|_{\infty}\:\leq\:\|\phi\|_{\infty}
\|\hat{P}\psi\|_1\:\leq\:2\|\phi\|_{\infty}$$
Therefore, 
$\hat{P}$ can be extended to a bounded operator on $\kom$. Clearly such an 
extension is also completely positive. In particular, it is strongly 
positive and so, for any $v\in{\cal H}$, $\|v\|\,=\,1$,
and any $\phi\in\tra$ we have
$$\|(\pef 
)v\|^2\;=\;<v,\,(\pef)^*(\pef)v>\:\leq\:<v,\,\hat{P}(\phi^*\phi)v>$$
$$=\;{\rm 
tr}P_v|\hat{P}(\phi^*\phi)\;=\;{\rm 
tr}\hat{P}(P_v)\phi^*\phi\:\leq\:\|\hat{P}
(P_v)\|_1\|\phi^*\phi\|_{\infty}\:\leq\:\|\phi\|_{\infty}^2$$
where 
$P_v=\,|v><v|$. However, $\hat{P}$ is a non-zero projection, hence 
$\|\hat{P}\|_{\infty,\infty}=\,1$. By duality, $\|\hat{P}\|_{1,1}=\,1$, 
too.\\
c) The dual operator $\hat{P}^*$ is a normal contraction on $\bha$. It 
is also a projection. Suppose $\phi,\psi\in\tra$. Then 
$${\rm 
tr}(\hat{P}^*\phi)\psi\;=\;{\rm tr}\phi\hat{P}(\psi)\;=\;{\rm 
tr}(\hat{P}\phi)\psi$$
Hence $\hat{P}^*|_{\tra}=\,\hat{P}$. However, $\tra$ 
is $\sigma$-weakly dense in $\bha$ so $\hat{P}^*$ is a normal extension of 
$\hat{P}$ onto $\bha$. We denote it by $\overline{P}$. The image of 
$\overline{P}$ equals to the $\sigma$-weak closure of ${\rm im}\hat{P}$ 
which coincides with the von Neumann algebra ${\cal M}$. $\Box$\\
Therefore, 
our task reduces to the description of the projection $\overline{P}$. By 
Prop.13 and 14 in [13] we know that ${\cal M}\,=\,\oplus_k{\cal 
M}_k\,=\,\oplus_k(\oplus_n\mkn)$,where $\mkn$ are type I factors. Let 
$E_{kn}$ denote the unit in $\mkn$ and let 
$\pna\,=\,E_{kn}\overline{P}(A)$. 
Then $\pkn$ is a projection onto $\mkn$ and 
$\sum_k\sum_n\pkn\,=\,\overline{P}$ since $\sum_k\sum_nE_{kn}\,=\,E$, the 
unit in $\cal M$. 
Hence
\be\overline{P}\;=\;\sum\limits_k\overline{P}_k\;
=\;\sum\limits_k(\sum\limits_n\pkn )\ee
where $\pkn$ are normal, norm one, and pairwise orthogonal projections, 
i.e. $\pkn\circ\overline{P}_{lm}\,=\,\overline{P}_{lm}\circ\pkn\,=\,0$ if 
$(kn)\neq(lm)$. Therefore, it suffices to determine the form of projections 
$\pkn$. Let us recall that each $\mkn$ has a minimal
projector $e_n$ of a 
finite dimension $r(k)$ for all $n$, where $r(k)$ is a subsequence of 
natural numbers. Let $N\,=\,N(k,\,n)$ be the degree of homogeneity (possibly 
infinite) of $\mkn$.\\
{\bf Theorem 4.2} {\it For} $A\in\bha$
\be\pna\;=\;\cun 
U(E_{kn}AE_{kn})U^*\ee
{\it where $\nkn\,=\,(\mkn')E_{kn}$ is the commutant 
of $\mkn$ in $\beh$, $\unn$ is the group of unitary operators
in $\nkn$, and 
$d\mu$ is a unique normalized Haar measure on} $\unn$.\\
{\bf Proof:} First, 
notice that for any projector $e$ in $\cal M$ the von Neumann algebras 
$e{\cal M}e$ and ${\cal M}_e$, where 
$${\cal M}_e\;=\;\{eA|_{{\rm 
range}\,e}:\;A\in{\cal M}\}\subset{\rm B}(e{\cal H})\}$$
are isomorphic. We 
use this identification in the proof. Let 
$V:E_{kn}{\cal H}\to\hikn\,=\,{\bf 
C}^N\otimes{\bf C}^{r(k)}$ be a unitary isomorphism. Here $\hikn$ denotes a 
Hilbert space being the direct sum of $N\,=\,N(k,\,n)$ copies of 
range$\,e_n$. $\mkn$ is isomorphic to the matrix algebra $M_{N\times 
N}(e_n\mkn e_n)$.
Because $e_n$ is minimal, so $e_n\mkn e_n=\,{\bf C}e_n$. 
Therefore, $\alpha(\mkn)\,=\,\bikn\otimes I_{r(k)}$, where 
$\alpha(A)\,=\,VAV^*$ for $A\in\beh$ and $I_{r(k)}$ is the identity 
$r(k)\times r(k)$ matrix. The projection $\alpha\circ\pkn 
|_{\beh}\circ\alpha^{-1}$ is the conditional expectation from 
$\bikn\otimes\mrk$ onto the first factor. Hence, for any 
$B\in\bikn\otimes\mrk$,
$$\alpha\circ\pkn 
|_{\beh}\circ\alpha^{-1}(B)\;=\;\crk({\bf 1}\otimes U)B({\bf 1}\otimes 
U^*)$$
and so 
$$\pna\;=\;\crk V^*({\bf 1}\otimes U)(VAV^*)({\bf 1}\otimes 
U^*)V^*$$
for any $A\in\beh$. However $V^*({\bf 1}\otimes U)V$ is a unitary 
operator in $\nkn$ and $\nkn$ is isomorphic to $\mrk$. Thus $\unn$ is 
isomorphic to $U(r(k))$. For a general $A\in\bha$ there is
$$\pna\;=\;\cun 
U(E_{kn}AE_{kn})U^*\quad\Box$$
Let us consider some particular cases of the 
projection $\overline{P}$.\\
{\bf Corollary 4.3} {\it If $r(1)\,=\,1$, then 
$\overline{P}_1$ is given by}
\be 
\overline{P}_1(A)\;=\;\sum\limits_n
\overline{P}_{1n}(A)\;=\;\sum\limits_nE_{1n}A 
E_{1n} \ee
{\it where $\{E_{1n}\}$ is a sequence (possibly finite) of 
pairwise orthogonal projectors of arbitrary dimensions. Let 
$J\,=\,\{(kn):\;N(k,\,n)\,=\,1\}$. Then for any $(kn)\in J$, ${\rm 
dim}E_{kn}<\infty$ and }
\be 
\sum\limits_J\overline{P}_{kn}(A)\;=\;\sum\limits_J{\rm 
tr}(E_{kn}A)\frac{E_{kn}}{{\rm dim}E_{kn}} \ee
{\it where $\{E_{kn}\}_J$ is a 
sequence of pairwise orthogonal finite dimensional projectors.}\\
{\bf 
Proof:} For $r(1)\,=\,1$ any $U\in{\cal U}({\cal N}_{1n})$ is of the form 
$U\,=\,e^{ia}E_{1n}$. Hence (6) and (7) implies (8). If $N(k,\,n)\,=\,1$, 
then $E_{kn}$ is a minimal projector in $\mkn$ with 
dim$E_{kn}\,=\,r(k)$. Therefore, $\pkn|_{\beh}$ is the conditional 
expectation onto ${\bf C}E_{kn}$ and formula (9) follows. $\Box$\\
Thus, if 
$r(1)=1$ we recover the Wan scheme [4], while for any $r(k)>1$ the minimal 
projectors in a corresponding coherent
subspace are of dimension $r(k)$ and 
so we meet the case of generalized rays. The restriction of the gauge group 
to subspace $E_{kn}{\cal H}$ is isomorphic to the unitary group ${\cal 
U}(r(k))$. Therefore, the whole gauge group equals to $\oplus_{r(k)>1}{\cal 
U}(r(k))$. In particular, if $N(k,\,n)=1$ for some $r$ and $k$ such that 
$r(k)>1$, then the restriction of algebra $\cal M$ to $E_{kn}{\cal H}$ 
consists only of numbers, i.e. ${\cal M}E_{kn}={\bf C}E_{kn}$. Hence we 
recover in formula (9) the coarse graining projection [16].

We meet another interesting case when $N(k,\,n)= r(k)$. 
Then, $E_{kn}{\cal H}$ is isomorphic to
${\bf C}^{r(k)}\otimes{\bf C}^{r(k)}$ and the 
restriction of $\cal M$ to $E_{kn}{\cal H}$ is isomorphic
to $M_{r(k)\times r(k)}\otimes I_{r(k)}$, that is effective observables 
act on the first factor, and so are isomorphic
with their commutant in B$(E_{kn}{\cal H})$. Such a situation was discussed 
by Giulini [17] in the context of the quantization of a system whose 
classical configuration space is not simply connected.

It is worth pointing out that all cases discussed above are of the
discrete type, that is all self-adjoint superselection operators have
discrete spectral decompositions.

Finally, we discuss the conservativeness of the 
induced superselection structure.\\
{\bf Proposition 4.4} {\it The induced superselection 
structure is conservative if and only if} $I\in{\cal M}$.\\
{\bf Proof:} $\Leftarrow$
If the identity operator belongs to $\cal M$, 
then $\overline{P}(I)=I$. Hence
${\rm tr}\hat{P}\rho\;=\;{\rm tr}\overline{P}(I)\rho\;
=\;{\rm tr}\rho$ for all $\rho\in\tra$.\\
$\Rightarrow$ Suppose now that 
tr$\hat{P}\rho={\rm tr}\rho$ for all $\rho\in\tra$. Let us assume on the
contrary that ${\cal M}$ does not contain $I$.
Then $\sum_{k,n}E_{kn}$ is a non-trivial projector. Let $E^{\bot}=\,I\,-\,
\sum_{k,n}E_{kn}$. For a state 
$\rho_0=E^{\bot}\rho_0 E^{\bot}$ we have that 
$\hat{P}(\rho_0)=0$ and so 
tr$\hat{P}(\rho_0)=0$, the contradiction. $\Box$\\
{\bf Corollary 4.5} {\it Suppose $T_t$ is 
relatively compact in the strong operator topology. Then the induced
superselection structure is conservative.}\\
{\bf Proof:} By Prop. 4.4 it suffices to show that 
$I\in{\cal M}$. Suppose on the contrary that $I$ does not 
belong to ${\cal M}$.
Then again for a state $\rho_0=\,E^{\bot}\rho_0E^{\bot}$ 
we have that $\hat{P}(\rho_0)=\,0$. However, by (5)
$$\lim\limits_{t\to\infty}\|T_t\rho_0\:-\:\hat{P}(T_t\rho_0)\|_1\;=\;
\lim\limits_{t\to\infty}\|T_t\rho_0\|_1\;=\;0$$
the contradiction, since $\|T_t\rho_0\|_1=\,{\rm tr}T_t\rho_0=\,1$ 
for all $t\geq 0$. $\Box$\\
Consequently, a strong superselection structure 
is always conservative and so the projection
$\overline{P}$ is a tr-compatible conditional 
expectation from $\bha$ onto $\cal M$. \\[4mm]
{\bf References}\\
$[1]$ Wick, G.C., Wightman, A.S., Wigner, E.P.: 
The intrinsic parity of elementary particles. Phys. Rev. {\bf 88},
101-105 (1952)\\
$[2]$ Wightman, A.S.: Superselection rules; old and new. Il Nuovo Cimento B {\bf 110}, 751-769 (1995)\\
$[3]$ Jauch, J.: System of observables in Quantum Mechanics. Helv. Phys. Acta {\bf 33}, 711-726 (1960)\\
$[4]$ Wan, K.: Superselection rules, quantum measurement, and Schr\"odinger's cat. Canadian J. Phys. {\bf 58},
976-982 (1980)\\
$[5]$ Messiah, A.M.L., Greenberg, O.W.: Symmetrization postulate and its experimental foundation. Phys. Rev.
{\bf 136B}, 248-267 (1964)\\
$[6]$ Jauch, J.M., Misra, B.: Sypersymmetries and essential observables. 
Helv. Phys. Acta {\bf 34}, 699-709 (1961)\\
$[7]$ Zurek, W.H.: Environment-induced superselection rules. Phys. Rev. D {\bf 26}, 1862-1880 (1982)\\ 
$[8]$ Joos, E., Zeh, H.D.: The emergence of classical properties through 
interaction with the environment. Z. Phys. B
{\bf 59}, 223-243 (1985)\\
$[9]$ Paz, J.P., Zurek, W.H.: Environment-induced decoherence, classicality, and consistency of quantum histories.
Phys. Rev. D {\bf 48}, 2728-2738 (1993)\\
$[10]$ Joos, E.: Decoherence through interaction with the environment. 
In: Giulini, D. et al. (eds.) Decoherence and the appearance of a classical world in quantum theory. 
Berlin: Springer 1996\\
$[11]$ Unruh, W.G., Zurek, W.H.: Reduction of a wave packet in Quantum Brownian motion. Phys. Rev. D {\bf 40},
1071-1094 (1989)\\
$[12]$ Twamley, J.: Phase-space decoherence: a comparison between consistent histories and environment-induced
superselection. Phys. Rev. D {\bf 48}, 5730-5745 (1993)\\
$[13]$ Olkiewicz, R.: Environment-induced superselection rules in Markovian regime. Commun. Math. Phys. (in press)\\
$[14]$ Davies, E.B.: Quantum Theory of Open Systems. Academic Press: London (1976)\\
$[15]$ Partovi, M.H.: Irreversibility, reduction and entropy increase in quantum measurement. Phys. Lett. A {\bf 137},
445-450 (1989)\\
$[16]$ Kupsch, J.: Open quantum systems.
In: Giulini, D. et al. (eds.) Decoherence and the appearance of a classical world in quantum theory. Berlin: Springer
1996\\
$[17]$ Giulini, D.: Quantum Mechanics on spaces with finite fundamental group. Helv. Phys. Acta {\bf 68}, 438-469
(1995)
\end{document}